\documentclass[12pt]{article}
\usepackage[left=1.0in,top=1.5in,right=1.0in,bottom=1.5in]{geometry}

\usepackage{setspace}
\usepackage{latexsym}
\usepackage{amsmath}
\usepackage{natbib}
\usepackage{bm}

\usepackage{graphicx}
\usepackage{amsfonts}
\usepackage{amssymb}
\usepackage{dcolumn}
\usepackage{amsthm}
\usepackage{amssymb}
\usepackage{float}
\usepackage{mbenotes}
\usepackage{threeparttable}
\usepackage{times}
\usepackage{natbib}
\usepackage[plain,noend,ruled]{algorithm2e}
\usepackage[hypertexnames=false,linktocpage=true, bookmarks=true]{hyperref}
\usepackage{caption}

\newtheorem{example}{Example}[section]
\newtheorem{remark}{Remark}[section]

\newcommand{\be}{\begin{equation}}
\newcommand{\en}{\end{equation}}
\newcommand{\bea}{\begin{eqnarray}}
\newcommand{\ena}{\end{eqnarray}}
\newcommand{\ba}{\begin{array}}

\newcommand{\ea}{\end{array}}

\usepackage{tkz-graph}  
\usepackage{calc}
\usepackage{tikz}
\usetikzlibrary{arrows}
\tikzstyle{vertex}=[circle, draw, inner sep=0pt, minimum size=6pt]

\begin{document}
\baselineskip .3in

\newcommand{\pr}{\mbox{pr}}
\newfont{\sss}{cmsy10 at 12pt}
\newcommand{\st}{Section }

\title{\textbf{\large Statistical Matching using Fractional Imputation  }}

\author{Jae-kwang Kim \and Emily Berg  \and Taesung Park}
\maketitle
\begin{abstract}
Statistical matching is a technique for integrating two or more data sets when information available for matching records for individual participants across data sets is incomplete. Statistical matching can be viewed as a missing data problem where a researcher wants to perform a joint analysis of variables that are never jointly observed. A conditional independence assumption is often used to create imputed data for statistical matching. 

We consider an alternative approach to statistical matching without using the conditional independence assumption. We apply parametric fractional imputation of \cite{kim11} to create imputed data using an instrumental variable assumption to identify the joint distribution. We also present variance estimators appropriate for the imputation procedure. We explain how the method applies directly to the analysis of data from split questionnaire designs and measurement error models.   
\end{abstract}

Key Words: Data combination, Data fusion,  Hot deck imputation,   Split questionnaire design, Measurement error model. 

\newpage

\section{Introduction}


Survey sampling is a scientific tool for making inference about the target population. However,  we often do not collect all the necessary information in a single survey, due to time and cost constraints.   
In this case, we wish to exploit, as much as possible, information already available from different data sources from the same target population.  Statistical matching, sometimes called data fusion \citep{baker89} or data combination \citep{ridder07},  aims to integrate two or more data sets when information available for matching records for individual participants across data sets is incomplete.  
\cite{dorazio06} and 
\cite{leulescu13} provide comprehensive overviews of the statistical matching techniques in survey sampling.

  Statistical matching can be viewed as a missing data problem where a researcher wants to perform a joint analysis of variables that are never jointly observed.   \cite{moriarity01} provide a theoretical framework for statistical matching under a multivariate normality assumption. \cite{rassler02} develops multiple imputation techniques for statistical matching with  pre-specified parameter values for non-identifiable parameters. \cite{lahiri05} address regression analysis with linked data.  \cite{ridder07} provide a rigorous treatment of the assumptions and approaches for statistical matching in the context of econometrics. 
  


\begin{table}[H]
\begin{center}
  \begin{tabular}{rcccc}
      \hline%
   \  \  \  \  \  \  \  \     &  \  \  $X$ \  \  & \  \  $Y_1$  \  \  &  \  \  $Y_2$ \   \ &  \  \    \   \   \   \   \   \ \\
       \cline{2-5}
\  \  \  \   \  \  \  \  \  \  \ Sample A   \  \  \  \  \ & o & o &  & \\
\  \  \  \  \  \  \   \  \  \  \  Sample B   \   \  \  \  \  & o &   &  o & \\
  \hline
\end{tabular}
\caption{A Simple data structure for matching}
\label{table1} 
  \end{center}
\end{table}

Statistical matching aims to construct fully augmented data files to perform statistically valid joint analyses. To simplify the setup, suppose that two surveys, Survey A and Survey B, contain partial information about the population. Suppose that we observe $x$ and $y_1$ from the Survey A sample and observe $x$ and $y_2$ from the Survey B sample.
Table \ref{table1} illustrates a simple data structure for matching.
If the Survey B sample (Sample B) is a subset of the Survey A sample (Sample A), then we can  apply record linkage techniques \citep{herzog07} to obtain values of $y_{1}$ for the survey B sample. However, in many cases, such perfect matching is not possible (for instance, because the samples may contain non-overlapping subsets), and we may rely on a probabilistic way of identifying the  ``statistical twins'' from the other sample. That is, we want to create $y_1$ for each element in sample B by finding the nearest neighbor from  Sample A. Nearest neighbor imputation has been discussed by many authors, including  \cite{chen01} and  \cite{beaumont09}, in the context of missing survey items.

 Finding the nearest neighbor is often based on ``how close'' they are in terms of $x$'s only. Thus, in many cases,  statistical matching is based on the assumption that $y_1$ and $y_2$ are independent, conditional on $x$.
That is,
\begin{equation}
 y_1 \perp y_2 \mid x.
 \label{cia}
 \end{equation}
 Assumption (\ref{cia}) is often referred to as the conditional independence (CI) assumption  and is heavily used in practice.

In this paper,  we consider an alternative approach that does not rely on the CI  assumption. Instead, we adopt an approach to statistical matching based on an instrumental variable, as discussed briefly in  \cite{ridder07}.  \cite{kimshao13} propose the fractional imputation method for statistical matching under an instrumental variable assumption. After we discuss the assumptions in Section 2, we review the fractional imputation methods in Section 3. Furthermore, we consider two extensions, one to split questionnaire designs  (in Section 4) and the other to measurement error models (in Section 5).  Results from two simulation studies are presented in Section 6. 

 \section{Basic Setup}

For simplicity of the presentation, we consider the setup of two independent surveys  from the same target population consisting of $N$ elements.
As discussed in \st 1, suppose that   Sample  A collects information only on $x$ and $y_1$ and  Sample B collects information only on $x$ and $y_2$.

 To illustrate the idea, suppose for now that $(x, y_1, y_2)$ are generated from a normal distribution such that
 $$ \begin{pmatrix}
 x \\ y_1 \\ y_2
 \end{pmatrix}
 \sim N \left[\begin{pmatrix}
 \mu_x \\ \mu_1 \\ \mu_2
 \end{pmatrix}, \begin{pmatrix}
 \sigma_{xx} & \sigma_{1x} & \sigma_{2x} \\
  & \sigma_{11} & \sigma_{12}  \\
 &  & \sigma_{22}
 \end{pmatrix} \right] .
 $$
  Clearly, under the data structure in Table \ref{table1}, the parameter $\sigma_{12}$ is not estimable from the samples. The conditional independence  assumption in (\ref{cia}) implies  that
$ \sigma_{12} = \sigma_{1x} \sigma_{2x}/\sigma_{xx} $
and $ \rho_{12} = \rho_{1x} \rho_{2x}$
That is, $\sigma_{12}$ is completely determined from other parameters, rather than estimated directly from the realized samples.

Synthetic data imputation under the conditional independence  assumption in this case can be implemented in two steps:
\begin{description}
\item{[Step 1]} Estimate $f(y_1 \mid x)$ from Sample A, and denote the estimate by $\hat{f}_a (y_1 \mid x)$.
\item{[Step 2]} For each element $i$ in Sample B, use the $x_i$ value to generate imputed value(s) of $y_1$ from $\hat{f}_a ( y_1 \mid x_i)$.
\end{description}
Since $y_1$ values are never observed in Sample B, synthetic values of $y_1$ are created for all elements in Sample B, leading to synthetic imputation.
\cite{haziza09} provides a nice review of literature on imputation methodology.   
 \cite{kimrao12} present a model-assisted approach to synthetic imputation when only $x$ is available in Sample B.
Such synthetic imputation completely ignores the observed information in $y_2$ from Sample B. 



\begin{figure}[H]
 \centering
 \begin{tikzpicture}[->,>=stealth',shorten >=1pt,auto,node distance=3cm,
  thick,main node/.style={circle, draw,font=\sffamily\Large\bfseries}]

  \node[main node] (Y1) {$Y_{1}$};
  \node[main node] (X2) [left of=Y1] {$X_{2}$};
  \node[main node] (Y2)[right of =Y1]{$Y_{2}$};
  \node[main node] (X1)[below right of= Y1]{$X_{1}$};
  
  \path[every node/.style={font=\sffamily\small}]
  (Y1) edge  node   [left] { }  (Y2)	
  (X2) edge node   [left]  { } (Y1) 
  (X1) edge node  [left]  { } (Y1)  
  (X1) edge node [left]  {  } (Y2);
  
\end{tikzpicture}
  \caption{Graphical illustration of the dependence structure for a model in which $x_{2}$ is an instrumental variable for $y_{1}$ and $x_{1}$ is an additional covariate in the models for $y_{2}$ and $y_{1}$.}
  \label{figgraph}
\end{figure}
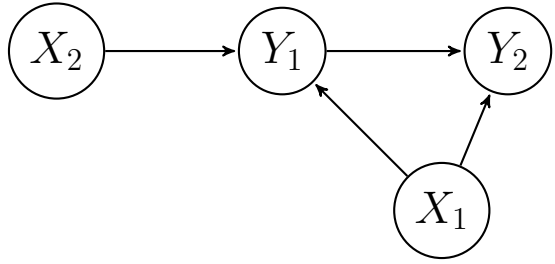

Statistical matching based on conditional independence  assumes  that $Cov(y_1, y_2 \mid x)=0$. Thus, the regression of $y_2$ on $x$ and $y_1$ using the imputed data from the above synthetic imputation will estimate a zero   regression coefficient for $y_1$.  That is, the estimate $\hat{\beta}_2$ for
    $$ \hat{y}_2 = \hat{\beta}_0 + \hat{\beta}_1 x + \hat{\beta}_2 y_1, $$
will estimate zero.  Such analyses can be misleading if CI does not hold. To explain why, we consider an omitted variable regression problem: 
\begin{eqnarray*}
y_1 &=& \beta_0^{(1)} + \beta_1^{(1)} x+ \beta_2^{(1)} z + e_1 \\
y_2 &=& \beta_0^{(2)} + \beta_1^{(2)} x + \beta_2^{(2)} z + e_2 
\end{eqnarray*}
where $z, e_1, e_2$ are independent and are not observed. Unless $\beta_2^{(1)}=\beta_2^{(2)}=0$,  the
latent variable $z$ is an unobservable confounding factor that explains why $Cov(y_1, y_2 \mid x) \neq 0$. Thus, the coefficient on $y_{1}$ in the population regression of $y_{2}$ on $x$ and $y_{1}$ is not zero, 
 
We consider an alternative approach which is not built on the conditional independence  assumption. First, assume that we can decompose $x$ as  $x=(x_1, x_2)$ such that
\begin{eqnarray*}
 (i) & &  f (y_2 \mid x_1, x_2, y_1 ) = f ( y_2 \mid x_1, y_1) \\
 (ii) & & f ( y_1 \mid x_1, x_2=a) \neq f( y_1 \mid x_1, x_2=b)
 \end{eqnarray*}
  for some $a \neq b$. Thus, $x_2$ is conditionally independent of $y_2$ given $x_1$ and $y_1$ but $x_2$ is correlated with $y_1$ given $x_1$.  Note that $x_1$ may be null or have a degenerate distribution, such as an intercept. The variable $x_2$ satisfying the above two conditions is often called an instrumental variable (IV) for $y_1$.  The directed acyclic graph in Figure~\ref{figgraph}  illustrates the dependence structure of a model with an instrumental variable. \cite{ridder07} used ``exclusion restrictions'' to describe the instrumental variable assumption.   One  example where the instrumental variable assumption is reasonable is repeated surveys. In the repeated survey, suppose that $y_t$ is the study variable at year $t$ and satisfies Markov property 
 $$ P( y_{t+1} \mid y_1, \cdots, y_{t} ) = P( y_{t+1} \mid y_{t} ),$$
where $P(y_{t})$ denotes a cumulative distribution function.  In this case, $y_{t-1}$ is an instrumental variable for $y_t$. In fact, any last observation of $y_s (s \le t)$ is the  instrumental variable for $y_t$.

  Under the instrumental variable assumption, one can use two-step regression to estimate the regression parameters of a linear model. The following example presents the basic ideas.

 \begin{example}
 Consider the two sample data structure in  Table \ref{table1}. We assume the following
  linear regression model:
 \begin{equation}
 y_{2i} = \beta_0 + \beta_1 x_{1i} + \beta_2  y_{1i} + e_i,
 \label{model9-3}
 \end{equation}
  where $e_i \sim (0, \sigma_e^2)$ and $e_i$ is independent of $(x_{1j}, x_{2j}, y_{1j})$ for all $i,j$.
 In this case, a consistent estimator of $\beta = (\beta_0, \beta_1, \beta_2)$ can be obtained by the two-stage least squares (2SLS) method as follows: \index{Two-stage least squares method} \index{2SLS method | \see{Two-stage least squares method}}
 \begin{enumerate}
 \item From sample $A$, fit the following ``working model'' for $y_1$
  \begin{equation}
   y_{1i} = \alpha_0 + \alpha_1 x_{1i} + \alpha_2 x_{2i} + u_i , \ \  \ u_i \sim (0, \sigma_u^2)
   \label{model9-4}
   \end{equation}
to obtain a consistent estimator of $\alpha=(\alpha_0, \alpha_1, \alpha_2)'$ defined by
 $$ \hat{\alpha} = (\hat{\alpha}_0, \hat{\alpha}_1, \hat{\alpha}_2)'= \left( X'X \right)^{-1} X' Y_1 $$
 where $X= \left[ X_0, X_1, X_2 \right]$ is a matrix whose $i$-th row is $(1, x_{1i}, x_{2i})$ and $Y_1$ is a  vector with $y_{1i}$ being the $i$-th component.
 \item A consistent estimator of $\beta=(\beta_0, \beta_1, \beta_2)'$ is obtained by the least squares method for the regression of $y_{2i}$ on $(1, x_{1i}, \hat{y}_{1i})$ where $\hat{y}_{1i}=\hat{\alpha}_0 + \hat{\alpha}_1 x_{1i} + \hat{\alpha}_2 x_{2i} $.
 \end{enumerate}
\label{example9.1}
 \end{example}

Asymptotic unbiasedness of the 2SLS estimator under the instrumental variable assumption  is discussed in Appendix A.  The 2SLS method is not directly applicable if the regression model (\ref{model9-3}) is nonlinear. Also, while the 2SLS method gives estimates of the regression parameters, 2SLS does not provide consistent estimators for more general parameters such as $\theta=Pr(y_2 <1 \mid  y_1 <3)$.  Stochastic imputation can provide a solution for estimating a more general class of parameters. We explain how to modify parametric fractional imputation of \cite{kim11} to address general purpose estimation in statistical matching problems.

\section{Fractional imputation}

We now describe the fractional imputation methods for statistical matching without using the CI assumption. The use of fractional imputation for statistical matching was originally presented in Chapter 9 of Kim and Shao (2013). 
To explain the idea, note that $y_1$ is missing in Sample B and  our goal is to generate $y_1$ from the conditional distribution of $y_1$ given the observations. That is, we wish to generate $y_1$ from
\begin{equation}
 f\left( y_1 \mid x, y_2 \right) \propto f \left( y_2 \mid x, y_1 \right) f \left( y_1 \mid x \right).
 \label{9-2}
 \end{equation}
To satisfy model identifiability, we may assume that $x_2$ is an IV for $y_1$. Under IV assumption, (\ref{9-2}) reduces to 
\begin{equation*}
 f\left( y_1 \mid x, y_2 \right) \propto f \left( y_2 \mid x_1, y_1 \right) f \left( y_1 \mid x \right).
 \end{equation*}
 
To generate $y_1$ from (\ref{9-2}), we can consider the following two-step imputation: 
\begin{enumerate}
\item Generate $y_1^*$ from $ \hat{f}_a \left( y_1 \mid x \right)$. 
\item  Accept  $y_1^*$  if $f\left( y_2 \mid x, y_1^* \right)$ is sufficiently large.
\end{enumerate}
Note that the first step is the usual method under the conditional independence  assumption. The second step incorporates the information in $y_2$. 
The determination of whether $f(y_{2}\,|\,x ,y_{1}^{*})$ is sufficiently large required for Step 2 is often made by applying a Markov Chain Monte Carlo (MCMC) method such as the Metropolis-Hastings algorithm \citep{chib1995understanding}. That is, let $y_1^{(t-1)}$ be the current value of $y_1$ in the Markov Chain. Then, 
we accept $y_1^*$ with probability  
$$ R(y_1^*, y_1^{(t-1)} ) = \min \left\{ 1, \frac{  f( y_2 \mid x, y_1^*)}{  f(y_2 \mid x, y_1^{(t-1)}) } \right\}. $$ 
Such algorithms can be computationally cumbersome because of slow convergence of the MCMC algorithm.

Parametric fractional imputation of \cite{kim11} enables generating imputed values in (\ref{9-2}) without requiring MCMC. The following EM algorithm by fractional imputation can be used:
\begin{enumerate}
\item For each $i \in B$,
generate $m$ imputed values of $y_{1i}$, denoted by $y_{1i}^{*(1)}, \cdots, y_{1i}^{*(m)}$,
from $ \hat{f}_a \left( y_1 \mid x_i \right)$, where $\hat{f}_a \left( y_1 \mid x \right)$ denotes the estimated density for the conditional distribution of $y_1$ given $x$ obtained from sample A.
\item Let $\hat{\theta}_t$ be the current parameter value of $\theta$ in
$f\left( y_2 \mid x, y_1 \right)$. For the $j$-th imputed value $y_{1i}^{*(j)}$, assign the fractional weight
$$ w_{ij(t)}^* \propto f\left( y_{2i} \mid x_{i}, y_{1i}^{*(j)}; \hat{\theta}_t \right)$$
such that  $ \sum_{j=1}^m w_{ij}^* = 1$.
\item Solve the fractionally imputed score equation for $\theta$
\begin{equation}
\sum_{i \in B} w_{ib} \sum_{j=1}^m  w_{ij(t)}^* S(\theta; x_{i},  y_{1i}^{*(j)} , y_{2i}) = 0 
\label{10}
\end{equation}
to obtain $\hat{\theta}_{t+1}$, where $S(\theta; x, y_1, y_2) = \partial \log f ( y_2 \mid x, y_1; \theta)/\partial \theta$, and $w_{ib}$ is the sampling weight of unit $i$ in Sample B.
\item  Go to step 2 and continue until convergence.
\end{enumerate}

In (\ref{10}), note that, for sufficiently large $m$, 
%
\begin{eqnarray*}
\sum_{j=1}^m  w_{ij(t)}^* S(\theta; x_{i},  y_{1i}^{*(j)} , y_{2i})  
& \cong & 
  \frac{ \int   S(\theta; x_{i},  y_1, y_{2i})  f( y_{2i} \mid x_{i}, y_{1i}^{*(j)}; \hat{\theta}_t ) \hat{f}_a ( y_1 \mid x_i ) dy_1 }{ \int  f( y_{2i} \mid x_{i}, y_{1i}^{*(j)}; \hat{\theta}_t ) \hat{f}_a ( y_1 \mid x_i ) dy_1  } \\
&=& E\left\{ S(\theta; x_{i},  Y_1, y_{2i}) \mid x_{i},  y_{2i} ; \hat{\theta}_t  \right\} .
\end{eqnarray*}
If $y_{i1}$ is categorical, then the fractional weight can be constructed by the conditional probability corresponding to the realized imputed value \citep{ibrahim90}. 
Step 2 is used to incorporate observed information of $y_{i2}$ in Sample B. Note that Step 1 is not repeated for each iteration. Only Step 2 and Step 3 are iterated until convergence. Because Step 1 is not iterated, convergence is guaranteed and the observed likelihood increases. See Theorem 2 of \cite{kim11}.

\begin{remark}
In Section 2, we introduce IV only because this is what it is typically done
in the literature to ensure identifiability.  The proposed method itself does not rely on this assumption.  To illustrate a situation where we can identify the model without introducing the IV assumption, 
suppose that the model is 
\begin{eqnarray*} 
y_2 &=&  \beta_0 + \beta_1 x + \beta_2 y_1 + e_2  \\
 y_1 &= & \alpha_0 + \alpha_1 x + e_1 
 \end{eqnarray*}
with $e_1 \sim N(0, x \sigma_1^2)$ and  $e_2 \mid e_1  \sim N(0, \sigma_2^2)$,  then 
$$ f( y_2 \mid x) = \int f( y_2 \mid x, y_1 ) f(y_1 \mid x) d y_1 $$
is also a normal distribution with mean $(\beta_0  + \beta_2 \alpha_0) +  (\beta_1   +  \beta_2 \alpha_1) x$ and variance $\sigma_2^2 + \beta_2^2 \sigma_1^2 x$. Under the data structure in Table 1, such a model is identified without assuming the IV assumption. \end{remark}

 Instead of generating $y_{1i}^{*(j)}$ from $\hat{f}_a( y_1 \mid x_{i} )$, we can consider a hot-deck fractional imputation (HDFI) method, where all the observed values of $y_{1i}$ in Sample A are used as  imputed values. In this case, the fractional weights in Step 2 are given by
    $$ w_{ij}^* (\hat{\theta}_t)  \propto w_{ij0}^*   f\left( y_{2i} \mid x_{i}, y_{1i}^{*(j)}; \hat{\theta}_t \right), $$
    where
\begin{equation}
 w_{ij0}^* = \frac{ \hat{f}_a (y_{1j} \mid x_{i} ) }{ \sum_{k \in A} w_{ka} \hat{f}_a (y_{1j} \mid x_{k} ) } .
 \label{fwgt0}
 \end{equation}
 The initial fractional weight $w_{ij0}^*$ in (\ref{fwgt0}) is computed by applying importance weighting with
$$ \hat{f}_a ( y_{1j}) = \int \hat{f}_a(y_{1j} \mid x ) \hat{f}_a (x) dx  \propto \sum_{i \in A} w_{ia} \hat{f}_{a} (y_{1j} \mid x_i )$$
as the proposal density for $y_{1j}$. 
 The M-step is the same as for parametric fractional imputation. See \cite{kimyang13} for more details on HDFI. In practice, we may use a single imputed value for each unit. In this case,  the fractional weights can be used as the selection probability in Probability-Proportional-to-Size (PPS) sampling of size $m=1$. 
 

For variance estimation, we can either use a linearization method or a resampling method. We first consider variance estimation for the maximum likelihood estimator (MLE) of $\theta$.  
If we use a parametric model ${f}(y_1 \mid x)= f( y_1 \mid x; {\theta}_1)$ and $f( y_2 \mid x, y_1; \theta_2)$,   the MLE of  $\theta=(\theta_1, \theta_2)$ is obtained by solving 
 \begin{equation}
 \left[ S_1(\theta_1) , \bar{S}_2(\theta_1, \theta_2) \right]= (0,0),
 \label{score}
\end{equation}
 where $S_1(\theta_1)= \sum_{i \in A} w_{ia} S_{i1} (\theta_1)$, $S_{i1}(\theta_1) = \partial \log f( y_{1i} \mid x_i; \theta_1)/ \partial \theta_1$  
 is the score function of $\theta_1$,    
 $$\bar{S}_2(\theta_1, \theta_2) = E\{ S_2(\theta_2) \mid X, Y_2; \theta_1, \theta_2 \}, $$
  $S_2(\theta_2)= \sum_{i \in B} w_{ib} S_{i2}(\theta_2) $,  and $S_{i2}(\theta_2) = \partial \log f (y_{2i} \mid x_i, y_{1i}; \theta_2)/ \partial \theta_2$ is the score function of $\theta_2$. Note that we can write $\bar{S}_2(\theta_1, \theta_2) =  \sum_{i \in B} w_{ib} E\{ S_{i2} (\theta_2) \mid x_i, y_{2i} ; \theta\}$. Thus, 
 \begin{eqnarray*}
 \frac{\partial }{\partial \theta_1'} \bar{S}_2 (\theta) &=&  \sum_{i \in B} w_{ib}
\frac{\partial }{\partial \theta_1'}
   \left[ \frac{\int S_{i2}(\theta_2)  f (y_{1} \mid x_i; \theta_1) f(y_{2i} \mid x_i, y_1; \theta_2) dy_1}{ 
   \int f (y_1 \mid x_i; \theta_1) f(y_{2i} \mid x_i, y_1; \theta_2) dy_1} \right]  \\
  &=&  \sum_{i \in B} w_{ib}  E\{ S_{i2}(\theta_2) S_{i1} (\theta_1)' \mid x_i, y_{2i}; \theta \} \\
  &- &  \sum_{i \in B} w_{ib}
   E\{ S_{i2}(\theta_2)  \mid x_i, y_{2i}; \theta \}E\{ S_{i1} (\theta_1)' \mid x_i, y_{2i}; \theta\}
  \end{eqnarray*}
  and
  \begin{eqnarray*}
 \frac{\partial }{\partial \theta_2'} \bar{S}_2 (\theta) &=&  \sum_{i \in B} w_{ib}
\frac{\partial }{\partial \theta_2'}
   \left[ \frac{\int S_{i2}(\theta_2)  f (y_{1} \mid x_i; \theta_1) f(y_{2i} \mid x_i, y_1; \theta_2) dy_1}{ \int f (y_1 \mid x_i; \theta_1) f(y_{2i} \mid x_i, y_1; \theta_2) dy_1} \right]  \\
   &=&   \sum_{i \in B} w_{ib} E\{ \frac{\partial }{ \partial \theta_2'} S_{i2}(\theta_2) \mid x_i, y_{2i}; \theta\}\\
  &+&  \sum_{i \in B} w_{ib}  E\{ S_{i2}(\theta_2) S_{i2} (\theta_2)' \mid x_i, y_{2i}; \theta\}\\
  &-&   \sum_{i \in B} w_{ib}E\{ S_{i2}(\theta_2)  \mid x_i, y_{2i}; \theta \}E\{ S_{2i} (\theta_2)' \mid x_i, y_{2i}; \theta \}.
  \end{eqnarray*}
Now,  $\partial \bar{S}_2 (\theta)/\partial \theta_1'$
 can be consistently estimated by
  \begin{equation}
  \hat{B}_{21} = \sum_{i \in B}w_{ib}  \sum_{j=1}^m w_{ij}^* S_{2ij}^*( \hat{\theta}_2) 
  \left\{  S_{1ij}^* (\hat{\theta}_1)- \bar{S}_{1i}^* (\hat{\theta}_1) \right\}' ,
  \label{9-bb}
  \end{equation}
  where $ S_{1ij}^* (\hat{\theta}_1)=  S_1(\hat{\theta}_1; x_i, y_{1i}^{*(j)})  $,  
  $ S_{2ij}^* (\hat{\theta}_2)=  S_2(\hat{\theta}_2; x_i, y_{1i}^{*(j)}, y_{2i})  $, and 
  $\bar{S}_{1i} ^*(\hat{\theta}_1) = \sum_{j=1}^m w_{ij}^* S_1(\hat{\theta}_1; x_i, y_{1i}^{*(j)})$. 
  Also, $\partial \bar{S}_2 (\theta)/ \partial \theta_2'$ can be consistently
  estimated by
  \begin{equation}
- \hat{I}_{22} = \sum_{i \in B} w_{ib} \sum_{j=1}^m w_{ij}^* \dot{S}_{2ij}^* (\hat{\theta}_2) -  \hat{B}_{22} 
\label{9-tau}
\end{equation}
where 
$$ \hat{B}_{22} 
= \sum_{i \in B} w_{ib} \sum_{j=1}^m w_{ij}^* S_{2ij}^* (\hat{\theta}_2) 
  \left\{ S_{2ij}^* (\hat{\theta}_2) - \bar{S}_{2i}^* (\hat{\theta}_2) \right\}' ,
$$
 $\dot{S}_{2ij}^* ({\theta}_2)= \partial S_2(\theta_2; x_i, y_{1i}^{*(j)}, y_{2i})/\partial \theta_2'$ and 
 $\bar{S}_{2i}^* (\theta_2) = \sum_{j=1}^m w_{ij}^* S_{2ij}^* (\theta_2)$.

Using a Taylor expansion with respect to $\theta_1$, 
\begin{eqnarray*}
\bar{S}_2 (\hat{\theta}_1, \theta_2 ) & \cong &  \bar{S}_2 ( {\theta}_1, \theta_2 ) 
 - E \left\{  \frac{\partial}{ \partial \theta_1'}\bar{S}_2 (\theta) \right\} \left[  E\left\{  \frac{\partial}{ \partial \theta_1'} S_1(\theta_1) \right\} \right]^{-1} S_1(\theta_1)
 \\
 &=&  \bar{S}_2 (\theta) +K  S_1(\theta_1),
\end{eqnarray*}
and  we can write
   $$
    V(\hat{\theta}_2) \doteq \left\{ E\left(\frac{\partial}{  \partial \theta_2' } \bar{S}_2 \right) \right\}^{-1}
    V\left\{ \bar{S}_2 (\theta ) + K S_1(\theta_1) \right\} \left\{ 
    E\left(\frac{\partial}{  \partial \theta_2' } \bar{S}_2 \right)    
    \right\}^{-1'}. $$
   Writing
   $$ \bar{S}_{2}(\theta ) = \sum_{i \in B} w_{ib} \bar{s}_{2i} (\theta ), $$
with $\bar{s}_{2i}(\theta) = E\{ S_{i2}(\theta_{2})\,|\, x_i, y_{2i}; \theta\}$, a consistent estimator of $V\left\{ \bar{S}_2 (\theta) \right\}$ can be obtained  by applying a design-consistent variance estimator to $\sum_{i\in B} w_{ib}\hat{s}_{2i}$ with
$ \hat{s}_{2i} = \sum_{j=1}^m w_{ij}^* S_{2ij}^* (\hat{\theta}_2) $. Under simple random sampling for Sample B, we have
$$ \hat{V}\left\{ \bar{S}_2 (\theta) \right\}
=n_B^{-2} \sum_{i \in B}  \hat{s}_{2i} \hat{s}_{2i}'. $$
Also,
$$  V\left\{  K S_1 (\theta_1) \right\}$$  is consistently estimated by
$$ \hat{V}_2 = \hat{K} \hat{V}(S_1) \hat{K}',   $$ 
where $\hat{K}=\hat{B}_{21} \hat{I}_{11}^{-1}$,  $\hat{B}_{21}$ is defined in (\ref{9-bb}), and $\hat{I}_{11}=  - \partial S_1 (\theta_1)/ \partial \theta_1'$ evaluated at $\theta_1=\hat{\theta}_1$. Since the two terms $\bar{S}_2 (\theta )$ and $S_1(\theta_1)$ are independent, the variance can be estimated by
$$ \hat{V}(\hat{\theta}) \doteq  \hat{I}_{22}^{-1}
\left[ \hat{V}\left\{ \bar{S}_2 (\theta ) \right\}  + \hat{V}_2\right] \hat{I}_{22}^{-1'} , $$
 where $\hat{I}_{22}$ is defined in (\ref{9-tau}). 
 
More generally, one may consider estimation of a parameter $\eta$ defined as a root of the census estimating equation $\sum_{i=1}^{N}U(\eta; x_i, y_{1i}, y_{2i})=0$.  Variance estimation of the FI estimator of $\eta$ computed from $\sum_{i \in B} w_{ib} \sum_{j=1}^m w_{ij}^{*}U(\eta; x_i, y_{1i}^{*(j)}, y_{2i})=0$ is discussed in Appendix B. 

\section{Split questionnaire survey design}

In \st 3, we consider the situation where Sample A and Sample B are two independent samples from the same target population. We now consider another situation of a split questionnaire design where the original sample $S$ is selected from a target population and then Sample A and Sample B are randomly chosen such that $A \cup B = S$ and $A \cap B = \phi$. We observe $(x,y_1)$ from Sample A and observe $(x, y_2)$ from Sample B. We are interested in creating fully augmented data with  observation  $(x, y_1, y_2)$ in $S$.

Such split questionnaire survey designs are gaining popularity because they reduce response burden \citep{raghu95, chipperfield09}. Split questionnaire designs have been investigated for the Consumer Expenditure survey  \citep{gonzalez08} and the National Assessment of Educational Progress (NAEP) survey in the US.  In applications of split-questionnaire designs, analysts may be interested in multiple parameters such as the mean of $y_{1}$ and the mean of $y_{2}$, in addition to the coefficient in the regression of $y_{2}$ on $y_{1}$.

To construct a fully augmented dataset in $S$, we still assume the instrumental variable assumption given in $(i)$ and $(ii)$ of \st 2. That is, we assume $x = (x_1, x_2)$, where $x_2$  satisfies  $f(y_{2}\,|\, x_{1}, x_{2}, y_{1}) = f(y_{2}\,|\, x_{1},y_{1})$ and $f(y_{1}\,|\, x_{1},x_{2} = a) \neq f(y_{1}\,|\,x_{1},x_{2} = b)$ for some $a\neq b$. One can use the sample data  for inference about the marginal distribution of $y_{1}$, the marginal distribution of $y_{2}$, and the conditional distribution of $y_{1}$ or $y_{2}$ given $x$. The instrumental variable assumption permits identification of the parameters defining the joint distribution of $y_1$ and $y_{2}$.  Estimators of parameters in the marginal distributions of $y_{1}$ and $y_{2}$ based on the fully imputed data set are more efficient than estimators based only on the sample data if $y_{1}$ and $y_{2}$ are correlated.  

In some split questionnaire designs (i.e. \cite{raghu95}), the sample design is constructed so that every pair of questions is assigned to some subsample. This restriction on the design permits inference for joint distributions. The instrumental variable assumption allows inference for joint distributions with more general designs where some pairs of questions (i.e., questions leading to responses $y_{2}$ and $y_{1}$) are never asked to the same individual. 

We consider a design where the original Sample $S$ is partitioned into two subsamples: $A$ and $B$. We assume that $x_{i}$ is observed for $i\in S$, $y_{1i}$ is collected for $i\in A$ and $y_{2i}$ is collected for $i\in B$. (For simplicity, we assume that no nonresponse occurs for either Sample $A$ or Sample $B$.) The probability of selection into $A$ or $B$ may depend on $x_{i}$ but can not depend on $y_{1i}$ or $y_{2i}$. As a consequence, the design used to select subsample $A$ or $B$ is non-informative for the specified model \citep[Chapter 6]{fuller09}.
 We let $w_{i}$ denote the sampling weight associated with the full sample $S$. We assume a procedure is available for estimating the variance of an estimator of the form $\hat{Y} = \sum_{i\in S} w_{i}y_{i}$, and we denote the variance estimator by $\hat{V}_{s}(\sum_{i\in S} w_{i}y_{i})$. 

A procedure for obtaining a fully imputed data set is as follows. First, use the procedure of \st 3 to obtain imputed values $\{y_{1i}^{*(j)}: i\in B, j = 1,\ldots, m\}$ and an estimate, $\hat{\theta}$, of the parameter in the distribution $f(y_{2}\,|\,y_{1},x_{1}; \theta)$.  The estimate $\hat{\theta}$ is obtained by solving 
\begin{equation}
 \sum_{i \in B} w_i \sum_{j=1}^m w_{ij}^* S_2(\theta; x_{1i}, y_{1i}^{*(j)}, y_{2i} ) = 0,
 \label{4-1}
 \end{equation}
where  $S_2(\theta; x_1, y_1, y_2) = \partial \log f(y_2 \mid y_1, x_1; \theta)/\partial \theta$.  
Given $\hat{\theta}$, generate imputed values $y_{2i}^{*(j)} \sim f(y_{2}\,|\,y_{1i},x_{1i}; \hat{\theta})$, for $i\in A$ and $j=1,\ldots, m$.  

Under the instrumental variable assumption, the parameter estimator $\hat{\theta}$ generated by solving (\ref{4-1}) is fully efficient in the sense that the imputed value of $y_{2i}$ for Sample A leads to no efficiency gain. To see this, note that the score equation using the imputed value of $y_{2i}$ is computed by 
\begin{equation}
 \sum_{i \in A} w_i m^{-1} \sum_{j=1}^m S_2(\theta; x_{1i}, y_{1i}, y_{2i}^{*(j)} ) 
+ \sum_{i \in B} w_i m^{-1}\sum_{j=1}^m w_{ij}^* S_2( \theta; x_{1i}, y_{1i}^*, y_{2i}) = 0. 
\label{4-2}
\end{equation}
Because $y_{2i}^{*(1)}, \cdots, y_{2i}^{*(m)}$ are generated from $f(y_{2}\,|\,y_{1i},x_{1i}; \hat{\theta})$,  
$$ p \lim_{m \rightarrow \infty}  \sum_{i \in A} w_i m^{-1} \sum_{j=1}^m S_2(\theta; x_{1i}, y_{1i}, y_{2i}^{*(j)} ) = \sum_{i \in A} w_i E\{ S_2(\theta; x_{1i}, y_{1i}, Y_2) \mid y_{1i}, x_{1i} ; \hat{\theta} \}.
$$
Thus, by the property of score function, the first term of (\ref{4-2}) evaluated at $\theta=\hat{\theta}$  is close to zero 
and the solution to (\ref{4-2}) is essentially the same as the solution to (\ref{4-1}). That is, there is no efficiency gain in using the imputed value of $y_{2i}$ in computing the MLE for $\theta$ in $f(y_{2} \mid y_1, x_1; \theta)$.

However, the imputed values of $y_{2i}$ can improve the efficiency of inferences for parameters in the joint distribution of $(y_{1i}, y_{2i})$. As a simple example, consider estimation of $\mu_{2}$, the marginal mean of $y_{2i}$. Under simple random sampling,  the imputed estimator of $\theta = \mu_2$ is
\begin{equation}
\hat{\theta}_{I,m} = \frac{1}{n} \left\{\sum_{i\in A}
\left( 
m^{-1}\sum_{j=1}^{m}y_{2i}^{*(j)}  \right)+ \sum_{i\in B}y_{2i}\right\}.
\end{equation}
  For sufficiently large $m$, we can write 
\begin{eqnarray*}
\hat{\theta}_{I,m} &=& \frac{1}{n} \left\{  \sum_{i \in A} \hat{y}_{2i} + \sum_{i \in B} y_{2i} \right\} \\
&=&  \frac{1}{n} \left\{  \sum_{i \in A} \left( \hat{\beta}_0 + \hat{\beta}_1 x_{1i} + \hat{\beta}_2 {y}_{1i} \right)+ \sum_{i \in B} y_{2i} \right\} , \end{eqnarray*}
where $(\hat{\beta}_0, \hat{\beta}_1, \hat{\beta}_2)$ satisfies 
$$ \sum_{i \in B} \left( y_{2i} - \hat{\beta}_0 - \hat{\beta}_1 x_{1i} - \hat{\beta}_2 \hat{y}_{1i} \right) = 0 $$
and $\hat{y}_{1i} = \hat{\alpha}_0 + \hat{\alpha}_1 x_{1i} + \hat{\alpha}_2 x_{2i} $ with $(\hat{\alpha}_0 , \hat{\alpha}_1, \hat{\alpha}_2)$ satisfying 
$$ \sum_{i \in A} \left( y_{1i} - \hat{\alpha}_0 - \hat{\alpha}_1 x_{1i}  - \hat{\alpha}_2 x_{2i} \right) = 0. $$

Under the regression model 
$$ y_{2i} = \beta_0 + \beta_1 x_{1i} + \beta_2 \hat{y}_{1i} + e_i $$
where $e_i \sim (0, \sigma_e^2)$, the variance of $\hat{\theta}_{I,m}$ is, for sufficiently large $m$,  
$$ V( \hat{\theta}_{I,m} ) =  \frac{1}{n}  V(y_2) + \left( \frac{1}{n_b} - \frac{1}{n} \right) V(e) 
$$
which is smaller than the variance of the direct estimator $\hat{\theta} = n_b^{-1} \sum_{i \in B} y_{2i}$.

\section{Measurement error models}

We now consider an application of statistical matching to the problem of measurement error models. Suppose that we are interested in the parameter $\theta$ in the conditional distribution $f(y\mid x; \theta)$. In the original sample, instead of observing $(x_i, y_i)$, we observe $(z_i, y_i)$, where $z_i$ is a contaminated version of $x_i$. Because inference for $\theta$ based on $(z_i, y_i)$ may be biased, additional information is needed. One common way to obtain additional information is to collect $(x_i, z_i)$ in an external calibration study. In this case, we observe $(x_i, z_i)$ in Sample A and $(z_i, y_i)$ in Sample B, where sample A is the calibration sample, and Sample B is the main sample. \cite{guo11} discuss an application of external calibration.

The external calibration framework can be expressed as a statistical matching problem.  Table \ref{table2} makes the connection between statistical matching and external calibration explicit.  The $(x_i, z_i, y_i)$ in the measurement error framework correspond to the $(y_{1i}, x_{2i}, y_{2i})$ in the setting of statistical matching. A straightforward extension of the measurement error model considered here incorporates additional covariates, such as the $x_{1i}$ of the statistical matching framework. 

\begin{table}
\begin{center}
\begin{tabular}{c|ccc}
\hline 
                                       & $z_i$ &   $x_i$  & $y_i$  \\ \hline \hline
Survey A (calibration study)   & o	&  o & \\ 
Survey B  (main study)   &     o    &    & o  \\ 
\hline 
\end{tabular}
\end{center}
\caption{Data structure for measurement error model}
\label{table2}
\end{table}

An instrumental variable assumption permits inference for $\theta$ based on data with the structure of Table 1. In the notation of the measurement error model, the instrumental variable assumption is 
\begin{eqnarray}
f(y_i\,|\,x_i,z_i) = f(y_i\,|\,x_i) \mbox{ and } f(z_i\,|\, x_i = a) \neq f(z_i\,|\,x_i=b),
\end{eqnarray}
for some $a\neq b$. The instrumental variable assumption may be judged reasonable in applications related to error in covariates because the subject-matter model of interest is $ f(y_i\,|\,x_i),$ and $z_i$ is a contaminated version of $x_i$ that contains no additional information about $y_i$ given $x_i$. 

 For fully parametric $f(y_i\,|\,x_i)$, $f(z_i\,|\,x_i)$ and $f(x_i)$, one can use parametric fractional imputation to execute the EM algorithm. This method requires evaluating the conditional expectation of the complete-data score function given the observed values. To evaluate the conditional expectation using fractional imputation, we first express the conditional distribution of $x$ given $(z,y)$  as,
\begin{eqnarray}
f \left( x \mid z, y \right) \propto  f \left( y \mid x \right) f ( x \mid z).
\end{eqnarray}
We let an estimator $\hat{f}_{a}(x_i\,|\,z_i)$ of $f(x_i\,|\,z_i)$ be available from the calibration sample (Sample A).  Implementation of the EM algorithm via fractional imputation involves the following steps:
\begin{enumerate}
\item For each $i \in B$, generate $x_i^{*(j)}$ from $\hat{f}_a ( x \mid z_i ) $, for $j = 1,\ldots, m$,
\item Compute the fractional weights
$$ w_{ij}^* \propto f ( y_i \mid x_i^{*(j)}; \hat{\theta}_{t} ) .$$
\item Update $\theta$ by solving
$$ \sum_{i \in B} w_{ib} \sum_{j=1}^m w_{ij}^* S(\theta; x_i^{*(j)} , y_i) = 0,$$
where $S(\theta; x_{i}^{*(j)}, y_{i}) = \partial \mbox{log}\{f \left( y \mid x; \theta \right)\}/ \partial \theta $. 
\item Go to Step 2 until convergence.
\end{enumerate}

The method above requires generating data from $f(x\,|\,z)$. For some nonlinear models or models with non-constant variances, simulating from the conditional distribution of $x$ given $z$ may require Monte Carlo methods such as accept-reject or Metropolis Hastings.  The simulation of Section 6.2 exemplifies a simulation in which the conditional distribution of  $x\,|\, z$ has no closed form expression. In this case, we may consider an  alternative approach, which may be computationally simpler. To describe this approach, let $h(x \mid z)$ be the ``working'' conditional distribution, such as the normal distribution, from which samples are easily generated. A special case of $h(x\mid z)$ is $f(x)$, the marginal density of $X$, which is used for selecting donors for HDFI.  
We assume that estimates $\hat{f}_{a}(x\mid z)$ and $\hat{h}_a(x \mid z)$ of  $f(x\mid z)$ and $h( x\mid z)$, respectively, are available from Sample A. Implementation of the EM algorithm via fractional imputation then involves the following steps:
\begin{enumerate}
\item For each $i \in B$, generate $x_i^{*(j)}$ from $\hat{h}_a ( x\mid z_i) $, for $j = 1,\ldots, m$,
\item Compute the fractional weights
\begin{eqnarray}\label{wformethod2}
w_{ij}^* \propto f ( y_i \mid x_i^{*(j)}; \hat{\theta}_{t} )\hat{f}_{a}(x_i^{*(j)}  \mid z_i)/ \hat{h}_a (x_i^{*(j)} \mid z_i).
\end{eqnarray}
\item Update $\theta$ by solving
$$ \sum_{i \in B} w_{ib} \sum_{j=1}^m w_{ij}^* S(\theta; x_i^{*(j)} , y_i) = 0 .$$
\item Go to Step 2 until convergence.
\end{enumerate}

\begin{remark}  
 Variance estimation is a straightforward application of the linearization method in \st 3. The hot-deck fractional imputation method described in \st 3 with fractional weights defined in (\ref{fwgt0}) also extends readily to the measurement error setting.  
For HDFI, the proposal distribution $\hat{h}_a(x \mid z)$ can be the empirical distribution with weights proportional to the sampling weights in Sample A. The imputed values are the $n_{A}$ values of $x_{i}$.  The weight $w_{ij}^{*}$ used for HDFI is 
\begin{eqnarray}\label{wformethod2hdfi}
w_{ij}^* \propto f ( y_i \mid x_i^{*(j)}; \hat{\theta}_{t} )\hat{f}_{a}(x_i^{*(j)} \mid z_i)/w_{ja},
\end{eqnarray}
where $ x_i^{*(j)} = x_{j}$ from sample $A$, and $w_{ja}$ is the associated sampling weight. 
\end{remark}
\section{Simulation study}

To test our theory, we present two limited simulation studies. The first simulation study considers the setup of combining two independent surveys of partial observation to obtain joint analysis. The second simulation study considers the setup of measurement error models with external calibration.

\subsection{Simulation One}
To compare the proposed methods with the existing methods, we generate
5,000 Monte Carlo samples of $(x_{i},y_{1i}, y_{2i})$ with size $n=400$, where 
$$\left(
\begin{array}{c}
y_{1i}\\x_i\\
\end{array}\right)
\sim N\left(\left[
\begin{array}{c}
2\\3\\
\end{array}\right],
\left[
\begin{array}{ccc}
1& 0.7\\
0.7 &1\\
\end{array}\right]\right), $$
\begin{equation}
y_{2i}=\beta_0+\beta_1y_{1i}+e_i, 
\label{16}
\end{equation}
 $e_i \sim N(0,\sigma^2)$,  and $\beta=(\beta_0,\beta_1,\sigma^2)=(1,1,1)$.    Note that, in this setup, we have $f(y_2 \mid x, y_1) = f(y_2 \mid y_1)$ and so the variable $x$ plays the role of the instrumental variable for $y_1$. 
 
Instead of observing $(x_i, y_{1i}, y_{2i})$ jointly, we assume that only  $(y_1, x)$   are observed in Sample A and only  $(y_2, x)$ are observed in Sample B, where Sample A is obtained by taking the first $n_a=400$ elements and Sample B is obtained by taking the remaining $n_b=400$ elements from the original sample. We are interested in estimating four parameters: 
three regression parameters $\beta_0, \beta_1, \sigma^2$ and  $\pi=P( y_1 < 2, y_2 < 3)$, the proportion of $y_1<2$ and $y_2<3$. Four methods are considered in estimating the parameters:
\begin{enumerate}
\item Full sample estimation (Full): Uses the complete observation of $(y_{1i}, y_{2i})$  in Sample B. 
\item Stochastic regression imputation (SRI): Use the regression of $y_1$ on $x$ from Sample A to obtain $(\hat{\alpha}_0, \hat{\alpha}_1, \hat{\sigma}_1^2)$, where the regression model is $y_1 = \alpha_0 + \alpha_1 x + e_1 $ with $e_1 \sim (0, \sigma_1^2)$.  For each $i \in B$, $m=10$ imputed values are generated by $y_{1i}^{*(j)} = \hat{\alpha}_0 + \hat{\alpha}_1 x_i + e_i^{*(j)}$ where $e_i^{*(j)} \sim N(0, \hat{\sigma}_1^2)$. 
\item Parametric fractional imputation (PFI) with $m=10$ using the instrumental variable assumption. 
\item Hot-deck fractional imputation (HDFI) with $m=10$ using the instrumental variable assumption.   
\end{enumerate}

Table \ref{table6.1} presents Monte Carlo means and Monte Carlo variances of the point estimators of the four parameters of interest. SRI shows large biases for all parameters considered because it is based on the conditional independence  assumption. Both PFI and HDFI provide nearly unbiased estimators for all parameters.  Estimators from HDFI are slightly more efficient than those from PFI because  the two-step procedure in HDFI uses the full set of respondents in the first step. The theoretical asymptotic variance of $\hat{\beta}_1$ computed from PFI is 
 \begin{eqnarray*}
  V\left(  \hat{\beta}_1 \right)
  & \doteq & \frac{1}{(0.7)^2} \frac{1}{400}  2 \left(1-\frac{0.7^2}{2} \right) +  \frac{1}{(0.7)^2} \frac{1}{400}  (1- 0.7^2) \doteq 0.0103     \end{eqnarray*}
which is consistent with the simulation result in Table \ref{table6.1}. 
 In addition to point estimation, we also compute variance estimators for PFI and HDFI methods. Variance estimators show small relative biases (less than 5\% in absolute values) for all parameters. Variance estimation results are not presented here for brevity.

\begin{table}
\begin{center}
\begin{tabular}{llrr}
 Parameter & Method & Mean & Variance \\    
     \hline \hline
      & Full &  1.00 & 0.0123 \\
       $\beta_0$ & SRI &  1.90  & 0.0869\\
     & PFI &  1.00 & 0.0472  \\
      & HDFI &  1.00 & 0.0465 \\  \hline 
     \vspace{1mm} 

           & Full &  1.00 & 0.00249 \\
  $\beta_1$    & SRI &  0.54  & 0.01648 \\
     & PFI &  1.00 & 0.01031 \\
    
     & HDFI &  1.00 & 0.01026 \\        \hline
     \vspace{1mm}
           & Full &  1.00 & 0.00482 \\
      $\sigma^2$   & SRI &   1.73 & 0.01657 \\
     & PFI &  0.99 & 0.02411 \\       
     & HDFI &  0.99 & 0.02270 \\  \hline
     \vspace{1mm} 
      & Full & 0.374 & 0.00058 \\  
      $\pi$    & SRI & 0.305 & 0.00255 \\
          & PFI & 0.375  & 0.00059  \\
       & HDFI & 0.375& 0.00057 \\ 
     \hline \hline
\end{tabular} 
\caption{Monte Carlo means and variances of point estimators from Simulation One. (SRI, stochastic regression imputation; PFI, parametric fractional imputation; HDFI; hot-deck fractional imputation)}
 \label{table6.1}
 \end{center}
\end{table}

The proposed method is based on the instrumental variable assumption.  To study the sensitivity of  the  proposed fractional imputation method, we performed an additional simulation study. Now, instead of generating $y_{2i}$ from (\ref{16}), we use 
\begin{equation}
y_{2i}=0.5+ y_{1i} + \rho(x_i - 3) +e_i, 
\label{17}
\end{equation}
where  $e_i \sim N(0, 1)$ and  $\rho$ can take non-zero values. We use three values of $\rho$, $\rho \in \{0, 0.1, 0.2\}$, in the sensitivity analysis and apply the same PFI and HDFI  procedure that is based on the assumption that  $x$ is an instrumental variable for $y_1$. Such assumption is satisfied for $\rho=0$, but it is weakly violated for $\rho=0.1 $ or $\rho=0.2$.  Using the fractionally imputed data in sample B, we estimated three parameters, $\theta_1=E(Y_1)$, $\theta_2$ is the slope for the simple regression of $y_2$ on $y_1$, and $\theta_3 = P( y_1 <2, y_2< 3)$, the proportion of $y_1<2$ and $y_2<3$. 
Table \ref{table6.2} presents Monte Carlo means and variances of the point estimators for three parameters under three different models.  In Table \ref{table6.2}, the absolute values of the difference between the fractionally imputed estimator and the full sample estimator increase  as the value of $\rho$ increases, which is expected as the instrumental variable assumption is more severely violated for larger values of $\rho$, but the  differences are relatively small for all cases. In particular, the estimator of $\theta_1$ is not affected by the departure from the instrumental variable assumption. This is because the imputation estimator under incorrect imputation model still provides unbiased estimator for the population mean as long as the regression imputation model contains an intercept term \citep{kimrao12}. 
Thus, this  limited sensitivity analysis suggests that the proposed method seems to provide comparable estimates when the instrumental variable assumption is weakly violated.

\begin{table}
\begin{center}
\begin{tabular}{lcrrr}
 Model  & Parameter & Method  & Mean  & Variance  \\    
     \hline \hline
 &      &  Full  & 2.00 & 0.00235   \\
  &    $\theta_1$ &  PFI &  2.00 & 0.00352 \\
   &   & FHDI &  2.00 & 0.00249 \\
  \cline{2-5}
  &     & Full &  1.00 &  0.00249\\
    $\rho=0$  &   $ \theta_2$  & PFI & 1.00 &  0.01031 \\
    &    & FHDI & 1.00 & 0.01026 \\
     \cline{2-5} 
      & & Full &  0.43 & 0.00061 \\ 
      &      $\theta_3$& PFI   & 0.43 &  0.00059\\
      &  & FHDI & 0.43 & 0.00057 \\
    \hline 
 &     & Full   & 2.00  &  0.00235  \\
     &    $\theta_1$& PFI & 2.00 & 0.00353 \\
      &  & FHDI & 02.00 & 0.00250 \\
       \cline{2-5}
  &    & Full & 1.07 & 0.00248 \\
    $\rho=0.1$    & $ \theta_2$  & PFI & 1.14 & 0.01091 \\     
        &   &    FHDI & 1.14 & 0.01081\\
        \cline{2-5}
   &     & Full & 0.44 & 0.00061 \\
      &  $ \theta_3$& PFI & 0.45 & 0.00062 \\        
      &  & FHDI & 0.45 & 0.00059 \\
      \hline 
 &   &  Full &  2.00 & 0.00235  \\
                        &    $\theta_1$ & PFI & 2.00 & 0.00353 \\
                         &   & FHDI & 2.00 & 0.00250 \\
                         \cline{2-5} 
   &    &  Full & 1.14  & 0.00250 \\
   $\rho=0.2$   &   $ \theta_2$& PFI  & 1.28 & 0.01115 \\
        &   & FHDI &  1.28  &  0.01102   \\
      \cline{2-5} 
      &  & Full & 0.44  & 0.00061 \\ 
        &  $\theta_3$& PFI  & 0.46 & 0.00066 \\
          &  & FHDI & 0.46 & 0.00062 \\
              \hline \hline
\end{tabular} 
\caption{Monte Carlo means and Monte Carlo variances  of the two point estimators for sensitivity analysis in Simulation One (Full,  full sample estimator;  PFI, parametric fractional imputation;  HDFI; hot-deck fractional imputation)}
 \label{table6.2}
 \end{center}
\end{table}

 \subsection{Simulation Two}

In the second simulation study, we consider a binary response variable $y_{i}$, where 
\begin{eqnarray}\label{modelfory}
y_{i}\sim\mbox{Bernoulli}(p_{i}), \\ \nonumber
\mbox{logit}(p_{i}) = \gamma_{0} + \gamma_{x}x_{i},
\end{eqnarray} 
and $x_{i}\sim N(\mu_{x},\sigma^{2}_{x})$. In the main sample, denoted by $B$, instead of observing $(x_{i}, y_{i})$, we observe $(z_{i}, y_{i})$, where
\begin{eqnarray}
z_{i} = \beta_{0} + \beta_{1}x_{i} + u_{i},
\end{eqnarray}
and $u_{i}\sim\mbox{N}(0,\sigma^{2}\,|\, x_{i}\,|\,^{2\alpha})$. We observe $(x_{i},z_{i})$, $i = 1,\ldots, n_{A}$ in a calibration sample, denoted by A.  For the simulation, $n_{A}=n_{B}=800$, $\gamma_{0} =1$, $\gamma_{x} = 1$, $\beta_{0} = 0$, $\beta_{1} = 1$, $\sigma^{2} = 0.25$, $\alpha = 0.4$, $\mu_{x} = 0$, and $\sigma^{2}_{x} = 1$. Primary interest is in estimation of $\gamma_{x}$ and testing the null hypothesis that $\gamma_{x} = 1$. The MC sample size is 1000. 

\hspace{.2 in } We compare the PFI and HDFI estimators of $\gamma_{x}$ to three other estimators. Because the conditional distribution of $x_{i}$ given $z_{i}$ is non-standard, we use the weights of (\ref{wformethod2}) and (\ref{wformethod2hdfi}) to implement PFI and HDFI, where the proposal distribution $\hat{h}_{a}(x_{i},\,|\, z_{i})$ is an estimate of the marginal distribution of $x_{i}$ based on the data from sample A. We consider the following five estimators: 

\begin{enumerate}
\item {\it PFI}: For PFI, the proposal distribution for genearating $x_{i}^{*(j)}$ is a normal distribution with mean $\hat{\mu}_{x}$ and variance $\hat{\sigma}^{2}_{x}$, where $\hat{\mu}_{x}$ and variance $\hat{\sigma}^{2}_{x}$ are the maximum liklihood estimates of $\mu_{x}$ and $\sigma^{2}_{x}$ based sample $A$.  The fractional weight defined in (\ref{wformethod2}) has the form, 
\begin{eqnarray}\label{specificwformesim}
w_{ij}^{*} \propto p_{i}^{y_{i}}(1-p_{i})^{1-y_{i}}\hat{f}_{a}(z_{i}\,|\,x_{i}),
\end{eqnarray}
where $p_{i}$ is the function of $(\gamma_{0},\gamma_{x})$ defined by (\ref{modelfory}), and $\hat{f}_{a}(z_{i}\,|\,x_{i})$ is the estimate of $f(z_{i}\,|\,x_{i})$ based on maximum likelihood estimation with the sample A data.  The imputation size $m = 800$. 

\item {\it HDFI}: For HDFI, instead of generating $x_{i}^{*(j)}$  from a normal distribution, the $\{x_{i}^{*(j)}: j=1,\ldots, 800\}$ are the 800 values of $x_{i}$ from sample $A$. 

\item {\it Naive}: A {\it naive} estimator is the estimator of the slope in the logistic regression of $y_{i}$ on $z_{i}$ for $i\in B$.  

\item {\it Bayes}: We use the approach of \cite{guo11} to define a Bayes estimator. The model for this simulation differs from the model of \cite{guo11} in that the response of interest is binary. We implement GIBBS sampling  with JAGS \citep{plummer03}, specifying diffuse proper prior distributions for the parameters of the model.  Letting
\begin{eqnarray*}
\theta_{1} = (\mbox{log}(\sigma^{2}_{x}), \mbox{log}(\sigma^{2}), \mu_{x},\beta_{0},\beta_{1},\gamma_{0},\gamma_{x}),
\end{eqnarray*}
we assume a priori that $\theta_{1} \sim \mbox{N}(0,10^{6}I_{7})$, where $I_{7}$ is a $7\times 7$ identity matrix, and the notation $N(0,V)$ denotes a normal distribution with mean 0 and covariance matrix $V$. The prior distribution for the power $\alpha$ is uniform on the interval $[-5,5]$. 

\hspace{.2 in } To evaluate convergence, we examine trace plots and potential scale reduction factors defined in \cite{gelman03} for 10 preliminary simulated data sets. We initiate three MCMC chains, each of length 1500  from random starting values and discard the first 500 iterations as burn-in. The potential scale reduction factors across the 10 simulated data sets range from 1.0 to 1.1, and the trace plots indicate that the chains mix well. To reduce computing time, we use 1000 iterations of a single chain for the main simulation, after discarding the first 500 for burn-in.   

\item A {\it Weighted Regression Calibration (WRC)} estimator. The WRC estimator is a modification of the weighted regression calibration estimator defined in \cite{guo11} for a binary response variable. The computation for the weighted regression calibration estimator involves the following steps:

\begin{enumerate}
\item[(i)] Using OLS, regress $x_{i}$ on $z_{i}$ for the calibration sample. 
\item[(ii)] Regress the logarithm of the squared residuals from step (i) on the logarithm of $z_{i}^{2}$ for the calibration sample. Let $\hat{\lambda}$ denote the estimated slope from the regression. 
\item[(iii)] Using WLS with weight $|z_{i}|^{2\hat{\lambda}}$, regress $x_{i}$ on $z_{i}$ for the calibration sample.  Let $\hat{\eta}_{0}$ and $\hat{\eta}_{1}$ be the estimated intercept and slope, respectively, from the WLS regression. 
\item[(iv)]  For each unit $i$ in the main sample, let $\hat{x}_{i} = \hat{\eta}_{0} + \hat{\eta}_{1}z_{i}$. 
\item[(v)]  The estimate of $(\gamma_{0},\gamma_{x})$ is obtained from the logistic regression of $y_{i}$ on $\hat{x}_{i}$ for $i$ in the main sample. 
\end{enumerate}

\end{enumerate}

\hspace{.2 in } Table~\ref{tab1me} contains the MC bias, variance, and MSE of the five estimators of $\gamma_{x}$. The naive estimator has a negative bias because $z_{i}$ is a contaminated version of $x_{i}$. The variance of the PFI estimator is modestly smaller than the variance of the HDFI estimator because the PFI estimator incorporates extra information through the parametric assumption about the distribution of $x_{i}$.  The PFI and HDFI estimators are superior to the Bayes and WRC estimators. 

\hspace{.2 in } We compute an estimate of the variance of the PFI and HDFI estimators of $\gamma_{x}$ using the variance expression based on the linear approximation.  We define the MC relative bias as the ratio of the difference between the MC mean of the variance estimator and the MC variance of the estimator to the MC variance of the estimator.  The MC relative biases of the variance estimators for PFI and HDFI are -0.0096 and -0.0093, respectively.

\begin{table}
\begin{center}
 \begin{tabular}{crrr}
\\
Method &  MC Bias   & MC Variance & MC MSE \\ 
\hline\hline
PFI    & 0.0239      & 0.0386          & 0.0392 \\ 
HDFI  & 0.0246      & 0.0387          & 0.0393 \\ 
Naive & -0.2241     & 0.0239          &  0.0742 \\ 
Bayes &  0.0406    & 0.0415           & 0.0432 \\ 
WRC &   0.112 & 0.0499   & 0.0625 \\ 
\hline\hline
\end{tabular}
\caption{Monte Carlo (MC) means, variances, and mean squred errors (MSE) of point estimators of $\gamma_{x}$ from Simulation Two. (PFI, parametric fractional imputation; HDFI, hot-deck fractional imputation; WRC, weighted regression calibration; MC, Monte Carlo; MSE, mean squared error)} 
  \label{tab1me}
\end{center}
\end{table}


\section{Concluding Remarks}

We approach statistical matching as a missing data problem and use PFI to obtain consistent estimators and corresponding variance estimators.  The imputation approach applies more generally than two stage least squares, which is restricted to estimation of regression coefficients in linear models. Rather than rely on the often unrealistic conditional independence assumption, the imputation procedure derives from an assumption that an instrumental variable is available. The measurement error framework of  Section 5 and Section 6.2, in which external calibration provides an independent measurement of the true covariate of interest, is a situation in which the study design may be judged to support the instrumental variable assumption.  Although the procedure is based on the instrumental variable assumption, the simulations of Section 6.1 show that the imputation method is robust to modest departures from the requirements of an instrumental variable.  


The proposed  methodology is applicable without the instrumental variable assumption, as long as the model is identified. If the model is not identifiable, then the EM algorithm for the proposed PFI method does not necessarily converge. In practice, one can treat the specified model identified if the EM sequence obtained from the specified model converges. The resulting analysis is consistent under the specified model.  This is one of the main  advantages of using the frequentist approach over Bayesian. In the Bayesian approach, it is possible to obtain the posterior values   even under non-identified models and the resulting analysis can be misleading.

Statistical matching can also be used to evaluate effects of multiple treatments in observational studies. By properly applying statistical matching techniques, we can create an augmented data file of potential outcomes so that causal inference can be investigated with the augmented data file \citep{morgan07}. Such extensions will be  presented elsewhere. 



\section*{Acknowledgment}

We thank Professor Yanyuan Ma, an anonymous referee and the AE for very constructive comments.
The research of the first author was partially supported by a grant from NSF (MMS-121339) and Brain Pool program (131S-1-3-0476) from Korean Federation of  Science and Technology Society. The research of the second author was supported by a Cooperative Agreement between the US Department of Agriculture Natural Resources Conservation Service and Iowa State University. 
The work of the third author was supported by the Bio-Synergy Research Project (2013M3A9C4078158) of the Ministry of Science, ICT and Future Planning through the National Research Foundation in Korea. 
\section*{Appendix}

\subsection*{A. Asymptotic unbiasedness of 2SLS estimator}

\renewcommand{\theequation}{A.\arabic{equation}}
\setcounter{equation}{0}

Assume that we observe $(y_1,x)$ in Sample A and observe $(y_2,x)$ in Sample B. 
To be more rigorous, we can write $(y_{1a}, x_a)$ to denote the observation  $(y_1,x)$ in Sample A. Also, we can write  $(y_{2b},x_b)$ to denote the observations in Sample B. In this case, the model can be written as 
\begin{eqnarray*}
y_{1a}&=&\phi_0 1_a +\phi_1 x_{1a}  + \phi_2 x_{2a} + e_{1a}\\
y_{2b}&=&\beta_0 1_b+ \beta_1 x_{1b} + \beta_2 y_{1b}+ e_{2b} 
\end{eqnarray*}
with $E(e_{1a} \mid x_a)= 0$ and $E(e_{2b} \mid x_b, y_{1b} )= 0$. Note that $y_{1b}$ is not observed from the sample. Instead, we use $\hat{y}_{1b}$ using the OLS estimate obtained from Sample A.

Writing $X_a=[ 1_a, x_a]$ and $X_b=[1_b, x_b]$, we have 
 $\hat{y}_{1b} =X_b ( X_a' X_a)^{-1} X_a' y_{1a}=X_b \hat{\phi}_a$. The 2SLS estimator of $\beta=(\beta_0, \beta_1, \beta_2)'$ is then 
$$ \hat{\beta}_{2SLS} = (Z_b' Z_b)^{-1} Z_b' y_{2b} $$
where $Z_b=[1_b, x_{1b},  \hat{y}_{1b}] $.  Thus, we have 
\begin{eqnarray}
\hat{\beta}_{2SLS} -\beta&=& (Z_b' Z_b)^{-1} Z_b' (y_{2b}- Z_b \beta ) \notag \\
&=& (Z_b'Z_b)^{-1} Z_b' \{ \beta_2 (y_{1b}- \hat{y}_{1b}) + e_{2b} \} . \label{a1}
\end{eqnarray}
We may write  
$$ 
y_{1b} = \phi_0 1_b +\phi_1 x_b +e_{1b} = X_b \phi + e_{1b} 
$$
where $E(e_{1b} | x_b)=0$. Since 
\begin{eqnarray*}
\hat{y}_{1b} &=& X_b (X_a' X_a)^{-1} X_a' y_{1a} \\
&=& X_b (X_a' X_a)^{-1} X_a'  ( X_a \phi + e_{1a}) \\
&=& X_b \phi + X_b (X_a' X_a)^{-1} X_a' e_{1a} ,
\end{eqnarray*}
we have 
$$ y_{1b} - \hat{y}_{1b} = e_{1b} - X_b (X_a' X_a)^{-1} X_a' e_{1a} $$
and (\ref{a1}) becomes 
\begin{equation}
\hat{\beta}_{2SLS} -\beta
= (Z_b'Z_b)^{-1} Z_b' \{ \beta_2 e_{1b} - \beta_2  X_b (X_a' X_a)^{-1} X_a' e_{1a}
+ e_{2b}
\}.
\label{a2}
\end{equation}

Assume that the two samples are independent. Thus, $E(e_{1b} \mid x_a, x_b, y_{1a})=0$. 
Also, $E\{ (Z_b' Z_b )^{-1} Z_b' e_{2b} \mid x_a, x_b, y_{1a}, y_{1b}\}=0$. Thus, 
\begin{eqnarray*}
 E\{ \hat{\beta}_{2SLS} -\beta \mid x_a, x_b,y_{1a}  \}& =&E\{ -\beta_2  (Z_b'Z_b)^{-1} Z_b'   X_b (X_a' X_a)^{-1} X_a' e_{1a} \mid x_a, x_b, y_{1a}  \} 
 \end{eqnarray*}
 and 
 \begin{eqnarray*}
  (Z_b'Z_b)^{-1} Z_b'   X_b (X_a' X_a)^{-1} X_a' e_{1a}  &=&  (Z_b'Z_b)^{-1} Z_b' \{   X_b (X_a' X_a)^{-1} X_a' (y_{1a} - X_a \phi ) \} \\
  &=& (Z_b'Z_b)^{-1} Z_b' 
 X_b ( \hat{\phi} _a- \phi)  . \end{eqnarray*}
 This term has zero expectation asymptotically because $n_b^{-1} Z_b' Z_b$ and  $n_b^{-1} Z_b' X_b$ are bounded in probability and $(\hat{\phi}_a-\phi)$ converges to zero.

\subsection*{B. Variance estimation }

\renewcommand{\theequation}{B.\arabic{equation}}
\setcounter{equation}{0}

Let the parameter of interest be defined by the solution to $U_N(\eta) = \sum_{i=1}^N U(\eta;  y_{1i}, y_{2i})=0$. We assume that $\partial U_N(\eta)/ \partial \theta=0$. Thus, parameter $\eta$ is priori independent of $\theta$ which is the parameter in the data-generating distribution of $(x,y_1,y_2)$.

Under the setup of \st 3, let $\hat{\theta}=(\hat{\theta}_1, \hat{\theta}_2)$ be the MLE of $\theta=(\theta_1, \theta_2)$ obtained by solving (\ref{score}).  Also, let $\hat{\eta}$ be the solution to $\bar{U}(\eta \mid \hat{\theta})=0$ where 
  $$ \bar{U}( \eta \mid \theta) = \sum_{i \in B} \sum_{j=1}^m  w_{ib} w_{ij}^* U( \eta; y_{1i}^{*(j)}, y_{2i} ), $$
and  $$ w_{ij}^* \propto f(y_{1i}^{*(j)} \mid x_i; \hat{\theta}_1) f( y_{2i} \mid y_{1i}^{*(j)} ; \hat{\theta}_2) / h(y_{1i}^{*(j)} \mid x_i) $$
with $\sum_{j=1}^m w_{ij}^*=1$. Here, $h(y_{1} \mid x)$ is the proposal distribution of generating imputed values of $y_1$ in the parametric fractional imputation. By introducing the proposal distribution $h$, we can safely ignore the dependence of imputed values $y_{1i}^{*(j)}$ on the estimated parameter value $\hat{\theta}_1$. 

By Taylor linearization, 
$$ \bar{U} ( \eta \mid \hat{\theta}) \cong \bar{U} (\eta \mid \theta ) + \left( \partial \bar{U} / \partial \theta_1'\right) ( \hat{\theta}_1 - \theta_1) +  \left( \partial \bar{U} / \partial \theta_2'\right) ( \hat{\theta}_2 - \theta_2)
$$
Note that 
$$ \hat{\theta}_1 - \theta_1  \cong \{ I_1(\theta_1) \}^{-1} S_1 ( \theta_1) $$
where $I_1 (\theta_1) = - \partial  S_1 (\theta_1)/ \partial \theta_1' $. Also, 
$$ \hat{\theta}_2 - \theta_2 \cong \left\{ -\frac{\partial}{ \partial \theta_2'} \bar{S}_2( \theta)  \right\}^{-1} \bar{S}_{2} (\theta) $$
 where 
 $$\bar{S}_2 ( \theta) = \sum_{i \in B} \sum_{j=1}^m w_{i} w_{ij}^*(\theta) S_2 (\theta_2 ; y_{1i}^{*(j)}, y_{2i} ) .$$
 Thus, we can establish 
 $$ \bar{U} ( \eta \mid \hat{\theta}) \cong \bar{U} (\eta \mid \theta ) + K_1 S_1(\theta_1) + K_2 \bar{S}_2 (\theta),
 $$
  where $K_1= D_{21} I_{11}^{-1} $ and $K_2 = D_{22} I_{22}^{-1} $ with $I_{11}= - E( 
  \partial S_1/ \partial \theta_1')$, $I_{22} = - E ( \partial \bar{S}_2 / \partial \theta_2' ) $, 
$D_{21} = E\{ U(\eta) S_1(\theta_1)' \}$ and $D_{22}=E\{ U(\eta) S_2(\theta_2)' \}$, we have
$$ V\{ \bar{U} (\eta \mid \hat{\theta}) \} = \tau^{-1} \left\{  V_1 + V_2 \right\} \tau^{-1'} $$
where $\tau = -E \{ \partial \bar{U} (\eta | \theta) /\partial \eta' \}$, 
$$ V_1= V\left\{  \sum_{i \in B} w_i ( \bar{u}_i^* + K_2 S_{2i}^* ) \right\} , $$
$\bar{u}_{i}^{*} = E[U(\hat{\eta}; y_{1i}, y_{2i})\,|\, y_{2i}; \hat{\theta}]$,  and  $ V_2 = V\{ K_1  \sum_{i \in A} w_i S_{1i} \} .$ 
A consistent estimator of each component can be developed similarly to \st 3.

\subsection*{C. Score Tests}

\renewcommand{\theequation}{A.\arabic{equation}}
\setcounter{equation}{0}

 In some applications related to measurement error, an analytical question of interest may be phrased in terms of a null hypothesis about the parameter $\theta$. Suppose that $\theta = (\theta_1, \theta_2)$, and the null hypothesis of interest is $H_{0}: \theta_2= \theta_{2,0}$ for a specified $\theta_{2,0}$. Hypotheses about functions of $\theta_1$ and $\theta_2$ can be expressed as a null hypothesis about a sub-vector of interest after appropriate reparametrization.  We define a score test using the approach of  \cite{rao98} and \cite{boos92}.

 Let
\begin{eqnarray}
U_{1i}(\theta_{1},\theta_{2}, \eta) = (U_{11i}(\theta_{1},\theta_{2},\eta), U_{12i}(\theta_{1},\theta_{2},\eta)),
\end{eqnarray}
where $U_{1ki}(\theta_{1},\theta_{2},\eta) = E[S_{1ki}(\theta_{1}, \theta_{2}, x_{i})\,|\,y_{i},z_{i},\eta]$ for $k=1,2$, 
\begin{eqnarray}
S_{1i}(\theta) = (S_{11i}(\theta_1, \theta_2, x_i), S_{12i}(\theta_{1}, \theta_{2}, x_{i})),
\end{eqnarray}
and $S_{1ki}$ is the vector of derivatives of the complete data log likelihood with respect to $\theta_k$. Under the null hypothesis, an estimator $\tilde{\theta}_{1}$ satisfies,
\begin{eqnarray}\label{u11theta1tilde}
U_{11}(\tilde{\theta}_{1}, \theta_{2},\eta) = \sum_{i\in B} w_{iB}U_{11i}(\tilde{\theta}_{1},\theta_{2,0},\hat{\eta})  = 0,
\end{eqnarray}
and we use parametric fractional imputation to solve (\ref{u11theta1tilde}). By a Taylor expansion,
\begin{eqnarray}\label{u11scoretaylor}
0 &=& U_{11}(\tilde{\theta}_{1}, \theta_{2,0},\hat{\eta}) \\ \nonumber
&\approx& U_{11}(\theta_{1},\theta_{2,0},\eta) + \tau_{1,11}(\tilde{\theta}_{1} - \theta_{1}) + \Delta_{1,\eta}(\hat{\eta} - \eta),
\end{eqnarray}
and
\begin{eqnarray}\label{u22scoretaylor}
U_{12}(\tilde{\theta}_{1},\theta_{2,0},\hat{\eta}) &\approx& U_{12}(\theta_{1},\theta_{2,0},\eta) + \tau_{1,21}(\tilde{\theta}_{1} - \theta_{1}) + \Delta_{2,\eta}(\hat{\eta} - \eta),
\end{eqnarray}
where $\tau_{1,k1}$ is the matrix of derivatives of $U_{1k}(\theta_{1},\theta_{2,0},\eta)$ with respect to $\theta_{1}$, and $\Delta_{k,\eta}$ is the matrix of derivatives of $U_{1k}(\theta_{1},\theta_{2,0},\eta)$ with respect to $\eta$. Solving (\ref{u11scoretaylor}) for $\tilde{\theta}_{1} - \theta_{1}$ and plugging the resulting expression into (\ref{u22scoretaylor}) gives,
\begin{eqnarray}
U_{12}(\tilde{\theta}_{1},\theta_{2,0},\hat{\eta}) &=& U_{12}(\theta_{1},\theta_{2,0},\eta)  - \tau_{1,21}\tau_{1,11}^{-1}\left\{U_{11}(\theta_{1},\theta_{2,0},\eta)\right\} \\ \nonumber && + (-\tau_{1,11}^{-1}\Delta_{1,\eta}, \Delta_{2,\eta})(\hat{\eta} - \eta).
\end{eqnarray}
An estimate of the variance of $U_{12}(\tilde{\theta},\theta_{2,0},\hat{\eta})$ is 
\begin{eqnarray}
\hat{V}_{s} = \hat{V}\{\sum_{i\in B} w_{iB}\hat{v}_{i}\} +  (-\hat{\tau}_{1,11}^{-1}\hat{\Delta}_{1,\eta}, \hat{\Delta}_{2,\eta})\hat{V}\{\hat{\eta}\}(-\hat{\tau}_{1,11}^{-1}\hat{\Delta}_{1,\eta}, \hat{\Delta}_{2,\eta})',
\end{eqnarray}
where
\begin{eqnarray}
\hat{v}_{i} = U_{12i}(\tilde{\theta},\theta_{2,0},\hat{\eta}) - \tau_{1,21}\tau_{1,11}^{-1}U_{11i}(\tilde{\theta}_{1},\theta_{2,0},\eta).
\end{eqnarray}
A size $\alpha$ score test of the null hypothesis, $H_{0}: \theta_{2} = \theta_{2,0}$ rejects if  $T(\theta_{2,0}) > \chi^{2}_{p}(1-\alpha)$, where
\begin{eqnarray}
T(\theta_{2,0}) = [U_{12}(\tilde{\theta}_{1},\theta_{2,0},\hat{\eta})]'\hat{V}_{s}^{-1}[U_{12}(\tilde{\theta}_{1},\theta_{2,0},\hat{\eta})],
\end{eqnarray}
and $\chi^{2}_{p}(\cdot)$ is the quantile function of a chi-squared distribution with $p$ degrees of freedom. A confidence region for $\theta_{2}$ with confidence level $1-\alpha$ is the set of $\theta_{2}$ with $T(\theta_{2,0} = \theta_{2}) < \chi^{2}_{p}(1-\alpha)$.


\end{document}